# ENHANCING SOFTWARE PRODUCT LINES WITH MACHINE LEARNING COMPONENTS

Luz-Viviana Cobaleda [1], Julián Carvajal [1], Paola Vallejo [2], Andrés López [1,3] and Raúl Mazo [3]

[1] Facultad de Ingeniería, Universidad de Antioquia, Medellín, Colombia.
[2] Escuela de Ciencias Aplicadas e Ingeniería, Universidad EAFIT, Medellín, Colombia.
[3] Lab-STICC, ENSTA, Brest, Francia.

## ABSTRACT

*Modern software systems increasingly integrate machine learning (ML) due to its advancements and ability to enhance data-driven decision-making. However, this integration introduces significant challenges for software engineering, especially in software product lines (SPLs), where managing variability and reuse becomes more complex with the inclusion of ML components. Although existing approaches have addressed variability management in SPLs and the integration of ML components in isolated systems, few have explored the intersection of both domains. Specifically, there is limited support for modeling and managing variability in SPLs that incorporate ML components. To bridge this gap, this article proposes a structured framework designed to extend Software Product Line engineering, facilitating the integration of ML components. It facilitates the design of SPLs with ML capabilities by enabling systematic modeling of variability and reuse. The proposal has been partially implemented with the VariaMos tool.*

## 1. INTRODUCTION

The rapid evolution of artificial intelligence (AI) over the last decade can be attributed to a convergence of key factors: enhanced computational power, the widespread availability of massive datasets, and the creation of more sophisticated algorithms. Consequently, AI has emerged as a transformative technological force, empowering software-intensive systems with novel capabilities in a wide range of domains [1], [2], [3], [4]. AI-based systems are essentially software systems whose functionalities are enabled by at least one AI component (e.g., for image and speech recognition or autonomous driving) [4]. However, incorporating AI components into software products introduces new software engineering challenges and amplifies existing ones. The situation becomes even more critical when these components are integrated not only into a single product but into a family of software products or a Software Product Line (SPL). Thus, the integration of Machine Learning (ML) components into SPLs introduces new dimensions of variability that traditional modeling techniques are not prepared to handle. This raises fundamental questions: How can an AI/ML component be modeled within an SPL? How can architects effectively integrate ML components into their SPLs? What information about the model is necessary to enable a successful SPL configuration process? The inability of current modeling approaches to address these questions reveals a significant research gap. Additionally, the integration of ML components into software systems introduces unique challenges that have given rise to the field of Software Engineering for AI (SE4AI). Recent literature has systematically identified the issues that emerge across the software lifecycle, impacting areas such as requirements engineering, architecture, testing, deployment, and maintenance [3], [4], [5], [6]. While these challenges are broad, this article focuses on those most relevant to the design of SPL.

Most research in SE4AI has focused, to date, on the challenges of integrating ML components into individual software systems. In the context of SPLs, where systematic reuse is the primary goal, these challenges not only persist but are magnified and transformed into variability management problems. For example, defining performance metrics for a product is an engineering challenge, but managing a catalog of components with different performance profiles to configure multiple products becomes a challenge in variability management. The literature that explicitly addresses this transformation of ML challenges in the SPL domain is notably scarce, representing a significant research gap. One of the main challenges documented for individual systems lies in requirements engineering, particularly in managing expectations. It has been reported that both customers [3] and development teams [6] often have limited knowledge of the actual capabilities and limitations of ML, leading to the establishment of unattainable requirements, such as requests for systems with no false positives or 100% accuracy [4]. This problem extends to the difficulty of translating business objectives into appropriate technical specifications, as the quantitative metrics used to characterize an ML model are often unintuitive to non-technical stakeholders [3]. Beyond the requirements, the dynamic nature of ML components introduces complex operational challenges. The literature highlights the emergence of new quality attributes, such as freshness and robustness, whose understanding is still fragmented [3], [4]. For instance, the freshness requirement addresses the performance degradation caused by phenomena like "concept drift" through continuous monitoring, which in turn necessitates defining both the tolerance for such degradation and the specific triggers for a model update [3]. Additionally, the management of these new attributes is complicated by the existence of inherent trade-offs, such as that between fairness and accuracy in a model [3], [4].

Although the challenges discussed are significant for individual software systems, their impact is amplified in the context of SPLs, where systematic reuse and variability management are paramount. The incorporation of ML components introduces additional variability issues, such as defining performance metrics at the product line level, aligning stakeholder understanding across multiple products, and specifying monitoring policies, that must be addressed not only for individual products but for product lines. Despite extensive research on AI-related software components, the current literature lacks approaches that explicitly consider the distinctive characteristics of these components within the context of SPLs [3], [4].

In this paper, we propose a framework for enhancing SPLs by enabling the seamless integration of ML components. Our main contribution is a specification-oriented approach that guides the integration of ML-based functionalities into SPLs. This approach addresses key aspects, including variability management, probabilistic feature modeling, ML component characterization, systematic ML component monitoring, systematic component replacement strategy, and derivation products with ML components. This strategy enables more systematic reuse, customization, and traceability of ML components across product configurations in the SPL context.

The remainder of the paper is structured as follows: Section 2 provides background on SPL engineering and ML components documentation. Section 3 details the proposed framework for designing SPLs with ML components and discusses the implications of this approach. Section 4 presents related work. Finally, Section 5 concludes the article and outlines directions for future work.

## 2. BACKGROUND

The design and development of SPLs rely on systematic approaches to manage variability and promote reuse across families of related software systems. To provide the necessary foundation for the proposed framework, this section outlines the core concepts of SPL engineering and the integration of ML components.

### 2.1. SPL and Variability Management

A SPL represents a systematic approach to developing families of related applications within a specific domain through strategic reuse of common assets [7]. This paradigm leverages shared components and systematic variability management to achieve significant reductions in development time and costs while improving product quality through the incorporation of proven, reusable artifacts.

Software Product Line Engineering (SPLE) operationalizes this approach through two fundamental processes, as presented in Figure 1: (1) Domain engineering, which establishes reusable assets and variability models, and (2) Application engineering, which derives specific products from these shared resources [7]. Variability—the capacity of a system to be adapted or configured for specific contexts—serves as the core mechanism enabling this systematic reuse across diverse product requirements.

**1) Domain engineering** establishes the foundation of reusable assets through two sequential phases. **A) Domain analysis** identifies and specifies SPL variability using formal models such as feature models [8], which define variation points, available alternatives, and constraint relationships. This phase encompasses: domain requirements definition to capture stakeholder needs and scope constraints, reference architecture specification aligned with domain requirements, and variability model quality assurance through systematic verification, diagnosis, and validation activities. **B) Domain implementation** transforms abstract specifications into concrete, reusable components. Key activities include requirements engineering for domain components, architectural design specification, domain component implementation, comprehensive unit testing, and explicit linkage between components and variability model elements. This phase produces the core asset base comprising domain components, architectural models, and associated test suites.

**2) Application engineering** derives specific products through the systematic configuration and instantiation of domain assets across two phases. **A) Configuration and customization management** captures customer-specific requirements and configures variability models accordingly, encompassing application requirements engineering, variability model configuration, application architecture definition, and component customization to meet specific product needs. **B) Derivation** constructs final products from configured domain assets through requirements engineering for the derivation process, assembly architecture definition, systematic product implementation from domain components, and comprehensive system integrity testing, including performance, validation, and audit verification.

This dual-process framework ensures systematic reuse while maintaining the flexibility necessary to address diverse product requirements within the target domain. The SPLE framework applies to various domains, including but not limited to education [9], agricultural systems [10], smart building [11], e-commerce [12], automotive manufacturing [13], and information systems [14].

As a representative example of an SPL, the virtual store SPL models a family of e-commerce platforms designed to support the online exchange of goods and services across diverse markets such as fashion, electronics, and digital content, this SPL captures a set of core functionalities common to most product instances, including a product catalog, shopping cart, payment module, and delivery system. These shared components are complemented by a range of variability points that allow customization according to specific business needs, such as authentication mechanisms, catalog presentation styles, supported payment gateways, search engines, content moderation tools, and user interface configurations. The SPL is designed to promote systematic reuse while enabling flexibility to address the functional and non-functional requirements of different virtual store deployments.

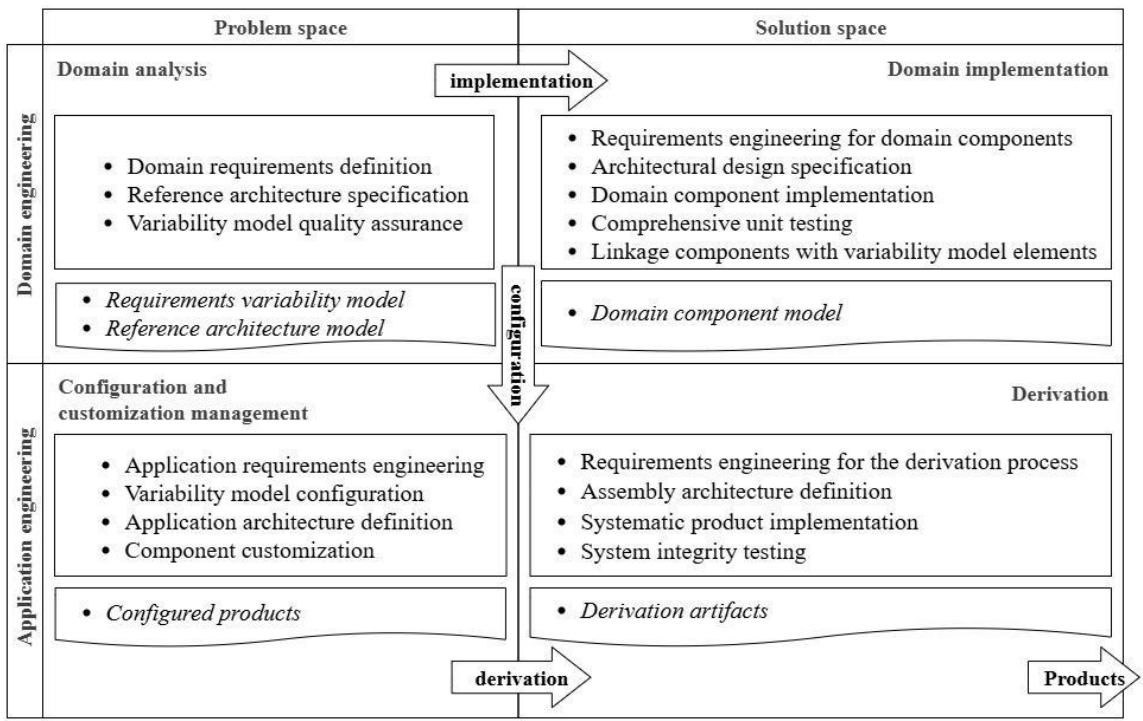

Figure 1. SPLE framework implemented in the VariaMos web tool, from [7]

## 2.2. ML Components

An **ML component** is a special type of software component that encapsulates ML models along with their associated data processing, inference logic, and system integration capabilities [6], [15]. These components constitute the main means of integrating ML capabilities into complex software systems, acting as a bridge between the underlying ML models and the overall system architecture. Component reuse is a foundational principle that enables the efficient development of multiple products from a shared, common core.

ML components can be deployed across various operational contexts depending on system requirements and architectural constraints. Common deployment patterns include: (1) embedded software libraries integrated directly within applications for low-latency scenarios; (2) standalone services accessible through REST or gRPC APIs for service-oriented architectures; and (3) containerized microservices within distributed cloud-native environments. The execution environment may range from local computational resources and edge devices to cloud-based infrastructures, each presenting distinct trade-offs in terms of latency, scalability, resource consumption, and operational complexity. The selection of an appropriate deployment strategy requires systematic evaluation of quality attributes, including performance characteristics, scalability requirements, maintainability constraints, and security considerations. These decisions must align with both functional requirements and non-functional system objectives, as the deployment choice significantly impacts the overall system behavior and operational characteristics. Given the critical role of ML models within these components, establishing a clear understanding of their behavior, limitations, and applicability conditions is essential for responsible reuse in systems.

In our representative example, the SPL for the virtual store also incorporates ML components, introducing a new dimension of intelligent behavior and adaptive functionality. These ML-based components include a **semantic search engine** that interprets the context of user queries, a **sentiment analysis module** applied to customer reviews, a **content moderation module** that filters out inappropriate content, and a **fraud detection module** to prevent and detect fraudulent activity. Each of these components can be configured in multiple ways depending on performance requirements, latency constraints, and human-in-the-loop considerations. For instance, the **content moderation module** may operate in a human-assisted mode or as a fully automated system, depending on the confidence thresholds applied to the underlying ML model. These components are encapsulated as reusable assets within the SPL, enabling developers to integrate advanced ML capabilities without retraining models from scratch for each product variant.

The selected components in the previous example are notable for both their practical applicability and their capacity to introduce significant variability dimensions—including model selection, performance thresholds, and human-in-the-loop configurations—that require explicit management within an SPL. However, in other domains, the set of selected components may differ. For instance, in the agricultural AI domain, relevant ML components for predictive modeling could include capabilities for carbon sequestration [16] or for emissions forecasting [17].

## 3. PROPOSAL

The integration of ML components into SPL represents a fundamental paradigm shift that challenges the traditional assumptions underlying systematic software reuse. While conventional SPL approaches have proven effective for deterministic software components with predictable behavior and stable interfaces [18], [19], ML components introduce unprecedented complexity through their inherent stochasticity, data dependency, continuous evolution requirements, and non-functional characteristics that defy traditional software engineering practices [3], [4].

This proposal presents a comprehensive framework that extends the foundational principles of product line engineering to accommodate the unique properties of ML components while preserving the economic and technical benefits that have made SPL a cornerstone of systematic software development. Our approach recognizes that ML components cannot be treated as "black boxes" within existing SPL methodologies; rather, they require a fundamental reconceptualization of domain modeling, component characterization, architectural design, and product derivation processes.

The framework proposed in this paper is currently being implemented in the VariaMos web tool (www.variamos.com) as part of an ongoing effort to operationalize and validate its practical applicability. It is organized around five interconnected phases that collectively address the complete lifecycle of ML-enhanced SPLs: **ML-aware domain analysis**, **Adaptive architecture design, ML-aware domainimplementation**, **Dynamic product configuration**, and **Product derivation and validation** of its resulting products. Each phase builds upon established SPL theory while introducing novel concepts and recommended practices specifically designed to handle the probabilistic nature, performance variability, and operational complexity inherent in ML systems. While VariaMos is a versatile, generic tool that accepts different domains, our examples pertain to the e-commerce domain.

### 3.1. ML-Aware Domain Analysis

The domain analysis phase requires significant adaptations when ML components are involved, particularly in feature modeling and architectural decision-making [7]. Traditional Boolean feature satisfaction proves inadequate for ML components whose capabilities vary across contexts and exhibit probabilistic behavior [6]. A key distinction of **ML-based features** lies in their inherent reliance on **training data properties**. The performance and functional capabilities of these features are susceptible to the characteristics of the training data, including its quality, representativeness, and intrinsic attributes. Additionally, implementing ML-based features can introduce risks associated with sensitive data, particularly regarding privacy, security, and information governance, due to the implications of data use and storage for model training and inference.

**Recommendation 1: Implement Probabilistic Feature Modeling.**

SPL engineers should extend conventional feature models to capture the uncertainty inherent in ML component capabilities[8]. Rather than relying on binary feature satisfaction, engineers should model features with quality distributions that reflect the variability in ML component performance.

**Practical Implementation:** For each feature that will be satisfied by an ML component, SPL engineers should identify it as an "*ML-based feature*" and define the following Feature Quality Profile:

```
FeatureQualityProfile = {
feature_id: String,
feature_type: type,
ml_component_id: String,
quality_distribution: {
      accuracy_range: [min_accuracy, max_accuracy],
      context_sensitivity: Map[Context, AccuracyLevel],
      confidence_intervals: Map[Scenario, ConfidenceRange]
}}
```

**E-commerce Example:** For a fraud detection feature in an online retail SPL, implementing recommendation 1, its Feature Quality Profile should look as follows:

```
FeatureQualityProfile = {
feature_id: "fraud_detection",
feature_type:ML-based,
ml_component_id: "fraud_detection_V1.0",
quality_distribution: {
      accuracy_range: [0.88, 0.95],
      context_sensitivity:{
            domestic_transactions_during_week:0.95,
            international_transactions_during_week: 0.88,
            domestic_transactions_during_weekend: 0.90,
            international_transactions_during_weekend: 0.75,
```

```
        transactions_from_suspicious_IP: 0.98,
        transactions_less_than_10_USD: 0.70},
    confidence_intervals: {
        high_confidence: [0.85, 1.0],
        medium_confidence: [0.70, 0.84],
        low_confidence: [0.0, 0.69]}}
}}
```

## 3.2. Adaptive Architecture Design

The reference architecture must explicitly address the dynamic and context-sensitive nature of ML components. ML models often evolve over time, depend on external data sources, and exhibit probabilistic behavior that affects system reliability and performance. Therefore, architectural decisions must incorporate design strategies that manage adaptability and traceability, ensure periodic updates, and maintain the long-term stability and performance of integrated ML functionalities. These strategies should align feature variability, model capabilities, and operational constraints, which is paramount for ensuring the robustness, adaptability, scalability, and maintainability of the ML-based SPL.

**Recommendation 2: Design ML-Aware Reference Architecture.**

The reference architecture must account for several key aspects.

- It must provide for a clear separation of concerns between the core SPL framework, the ML model development cycle, the deployment pipeline, and model monitoring.
- It must support various deployment strategies, including on-device (edge computing), on-premises, or cloud-based, depending on the specific product requirements and constraints.
- It must ensure data privacy, security, and compliance, while facilitating seamless integration with robust ML engineering practices, such as MLOps.

**Practical Implementation:** SPL engineers should be able to:
- Use microservice-based architecture, where ML components are deployed as decoupled services accessible through well-defined APIs.
- Use of containerization (e.g., Docker) to package models and their dependencies, ensuring environmental consistency and portability.

## 3.3. ML-Aware Domain Implementation

The domain implementation phase requires a structured approach to documenting, versioning, and managing ML components. This approach should be complemented by a formal monitoring process that can detect performance degradation and automatically trigger component replacement procedures. Effectively characterizing and selecting suitable ML components is essential to understanding their capabilities, limitations, and performance profiles. This enables successful integration and reduces associated risks. The monitoring system is designed to address the dynamic and non-deterministic nature of ML components by identifying potential degradation in the production environment and issuing alerts. Additionally, careful consideration is required for some aspects. For example, orchestrating ML components across products involves managing dependencies, activation conditions, and contextual adaptation. Furthermore, replacing ML components systematically requires mechanisms to evaluate, decouple, and reintegrate new versions with minimal disruption.

**Recommendation 3: Adopt Intelligent Component Characterization.**

To ensure the precise and systematic characterization of pre-trained ML components, it is proposed that **Model Cards** be mandatorily adopted. **Model Cards**, introduced by Mitchell [20] and further extended by Toma [21], provide a standardized framework for documenting ML models in a transparent and structured manner. This approach recommends customizing specific sections of the standard Model Card, such as Model Details, Intended Use, SPL reusability Profile, Model Usage, Operational Requirements, Performance Metrics, and Caveats. These cards are tailored for domain experts who, while not data scientists, are responsible for selecting and integrating third-party components.

**Practical Implementation:** For each ML component in the SPL, a standardized model card is proposed, capturing the following essential information:

```
ModelCard= {
model_details: {
      model_id: String,
      version: ModelVersion,
      developed_by: String,
      model_type: MLModelType,
      license: LicenseSpecification
}, intended_use: {
      primary_use: String,
      out-of-scope_use: String
}, spl_reusability_profile: {
      supported_domains: Set[Domain],
      integration_complexity: String, (ej. "Low")
}, model_usage: {
      api_endpoint: String,
      deployment_guidance: String
}, performance_metrics: Map[clave, valor],
   operational_requirements: {
          cpu: CPUSpecification, ram: RAMSize,
          gpu: String, notes: String
   },
   caveats: [String]
}
```

The SPL-aware Model Card specification defines the essential attributes for characterizing an ML component. The purpose and content of each key attribute are detailed below:

- **model_details**: Provides technical specifications—covering developer information, version control, model architecture, training methodology, and licensing terms that define commercial use rights, current license type, and redistribution permissions.

    - **model_id**: A unique identifier for the model, such as its name in a public repository.
    - **version**: The specific version of the model, following semantic versioning where possible, to track changes and dependencies.
    - **developed_by**: The organization, team, or individual responsible for the model's development.
    - **model_type**: Specifies the model's task category (e.g., Text Classification, Object Detection), informing its functional role.
    - **license**: The legal specification governing the use, modification, and distribution of the model, crucial for commercial product derivation.

- **intended_use**: Defines appropriate use cases, target applications, and intended user populations by outlining usage scenarios, specifying primary and out-of-scope applications, detailing the model's adaptability, and highlighting its limitations and potential biases.

    - **primary_use**: A concise description of the model's main purpose and the scenarios where it is designed to be applied (e.g., real-time fraud detection).
    - **out-of-scope_use**: Explicitly states the limitations and use cases for which the model has not been designed or validated, preventing misuse.

- **spl_reusability_profile**: A section dedicated to evaluating the ML component's fitness as a reusable asset within the SPL context. This is a key input for variability modeling.

    - **supported_domains**: A set of application domains where the model has demonstrated reliable performance, highlighting potential domain biases.
    - **integration_complexity**: A categorical rating (e.g., "Low", "Medium", "High") that estimates the engineering effort needed to integrate the component, based on its dependencies and API.

- **Model_usage**: Offers guidance on model consumption through various interfaces (e.g., UI, API) and outlines its compatibility with different deployment platforms and operating systems.It also provides guidance on optimizing performance and outlines deployment strategies for different environments, including local setups and cloud platforms.

    - **api_endpoint**: The URL or interface for sending inference requests.
    - **deployment_guidance**: A summary of instructions and best practices for deploying the model in different environments (e.g., cloud, edge).

- **performance_metrics**: Comprehensive performance evaluation including accuracy measures, uncertainty quantification, and decision thresholds.

- **operational_requirements**: Provides system requirements and hardware recommendations to help users prepare for deploying or fine-tuning the model in their computing environment.

    - **cpu**: The recommended minimum specification for the CPU. This is critical for overall system performance and serves as the primary compute resource when no GPU is used.
    - **ram**: The recommended minimum system RAM. This memory is required to hold the operating system, host application, model dependencies, and the model itself before being loaded into specialized hardware.
    - **gpu**: Specify whether a GPU is required, as well as its minimum specifications.
    - **notes**: Provides additional qualitative context or performance tips.

- **caveats and recommendations**: Presents caveats and recommendations by assessing potential societal impacts, fairness considerations, and bias mitigation strategies, while also outlining behavioral limitations related to "Not Safe For Work" (NSFW) content such as explicit material, violence, or hate speech.

**E-commerce Example:** The Model Card of the ML component of sentiment analysis would be:

```
ModelCard= {
model_details: {
```

```
        model_id: "tc_001",
        version: 2,
        developed_by: "Hugging Face",
        model_type: Text Classification,
        license: Apache-2.0,
},
intended_use: {
        primary_use: "The model can be used for topic classification",
        out-of-scope_use: "The model was not trained to be factual or true
        representations of people"
},
model_usage: {
        api_endpoint: https://plapplication.com/sentimentAnalisys1/predict,
        deployment_guidance:   http://huggingface.co/distilbert/distilbert-base-
        uncased
},
spl_reusability_profile: {
        supported_domains: ["Movies", "Series", "Music", "Products"],
        integration_complexity: "Low"
},
performance_metrics: ["Accuracy": 91.3],
operational_requirements: {
        cpu: 2+ CPU Cores, ram: 4GB, gpu: "Optional",
        notes: "Although the GPU is optional, its inclusion can significantly
        improve performance for some scenarios"
},
caveats_recommendations: ["The model is vulnerable to producing biased
predictions affecting underrepresented groups. For instance, when evaluating
sentences such as "This film was filmed in COUNTRY," the model assigns
drastically different probabilities to the positive label based on the country
mentioned (e.g., a 0.89 probability for France versus 0.08 for Afghanistan)."]
}
```

This information empowers the SPL architect to make a reasoned configuration decision: either accept a component with known limitations and plan for specific monitoring, or select an alternative component whose characteristics are better aligned with the product being built. In addition, the systematic adoption of Model Cards represents a crucial step toward responsible ML deployment by enhancing transparency around model behavior and operational boundaries. By standardizing both technical and ethical documentation practices, Model Cards enable stakeholders to evaluate and compare models using multidimensional criteria that extend beyond traditional performance metrics to encompass fairness, inclusivity, and equity considerations.

**Recommendation 4: Implement Systematic ML Component Monitoring.**

Given the inherently non-deterministic and data-dependent behavior of ML components, SPL engineers must design robust monitoring mechanisms capable of detecting performance degradation [22]. Operating at runtime, these mechanisms should continuously observe both model performance and business-critical signals, while being seamlessly integrated with drift detection and alerting processes to ensure resilient and self-adaptive system behavior.

**Practical Implementation:** To effectively implement this recommendation, SPL engineers should define a dedicated ML monitoring component for each domain component that incorporates ML capabilities. This component must specify the following attributes:

```
MLComponentMonitor: {
    component_id: String,
    monitoring_configuration: {
      metrics: Set[MonitoringMetric],
      frequency: TemporalSpecification,
      data_collection_strategy: DataCollectionApproach,
      baseline_establishment: BaselineDefinition
    },
    threshold_definitions: {
      performance_thresholds: Map[Metric, ThresholdSpec],
      drift_detection_thresholds: Map[DriftType, ThresholdSpec],
      business_impact_thresholds: Map[BusinessMetric, ThresholdSpec]
    },
    intervention_strategies: {
      alert_procedures: AlertSpecification
    }
}
```

The specification defines the structural requirements needed to establish consistent, interpretable, and actionable monitoring configurations. The key attributes of the monitoring specification are detailed below:

- **component_id**: Unique identifier of the monitored ML component. Used to record events, logs, and monitoring metrics.

- **monitoring_configuration**: Parameters that define what, how, and when monitoring is performed.

    - **metrics**: Set of key metrics for monitoring model performance. These metrics depend on the type of ML model (e.g., classification [*F1Score, AUC, Accuracy*], regression [*RMSE, MAE*], recommendation [*Precision, Recall*]).
    - **frequency**: Frequency at which the model's status is evaluated. It may depend on the traffic rate or importance of the model (e.g., *Hourly*: useful for high-volume production; *Daily*: balanced for general use; *EveryBatch*: suitable for batch systems; *RealTime*: when online processing is used).
    - **data_collection_strategy**: Method for collecting input data (for comparison and evaluation), predictions, and actual labels (if available) (e.g., *StreamingLogs*: continuous online capture. (e.g., *BatchLogs*: data collected in intervals; *ShadowDeployment*: evaluates without exposing to the user; *MiddlewareCapture*: collects from a proxy or wrapper).
    - **baseline_establishment**: Reference against which current metrics are compared. It can be a previous version or a historical average. (e.g., *StaticThresholds*: defined by experts; *PrelaunchModelBaseline*: based on offline evaluation; *Rolling7DayAverage*: adaptive and dynamic).

- **threshold_definitions**: Set of thresholds that trigger alerts.

    - **performance_thresholds**: Thresholds over key model quality metrics.
    - **drift_detection_thresholds**: Statistical thresholds for detecting changes in the distribution (data drift, concept drift, prediction drift, etc).
    - **business_impact_thresholds**: Business metrics that may be impacted by the model, such as CTR, revenue, and churn.

- **intervention_strategies**: Defines actions to take if an anomaly or system degradation is detected.

    ○ **alert_procedures**: Specification of the channel and form of alert to the responsible team (e.g., SendMailToMLTeam, PushToPagerDuty).

**E-commerce Example:** In an online retail SPL, the sentiment analysis component can be continuously monitored to detect potential performance degradation, drift, or business impact issues. The following configuration illustrates how a monitoring component can be defined to track relevant metrics and trigger intervention strategies when necessary.

```
MLComponentMonitor: {
  component_id: "tc_001",
  monitoring_configuration: {
    metrics: ["Precision", "Recall"],
    frequency: "Daily",
    data_collection_strategy: "StreamingLogs",
    baseline_establishment: "Rolling7DayAverage"
  },
  threshold_definitions: {
    performance_thresholds: {
        Precision: {min: 0.94, critical: 0.89, window: "24h"},
        Recall: {min: 0.87, critical: 0.82, window: "24h"}
    },
    drift_detection_thresholds: {
      DataDrift: {
        metric: "KL-Divergence", warning: 0.04, critical: 0.08, window: "7d"
      },
      ConceptDrift: {
        metric: "JS-Divergence", warning: 0.03, critical: 0.07, window: "7d"
      }
    },
    business_impact_thresholds: {
        misclassified_negative_reviews: {
        warning: 200, critical: 400, window: "24h"}
    }
  },
  intervention_strategies: {
    alert_procedures: {
      warning_level: "SendMailToMLTeam", critical_level: "PushToPagerDuty"
    }
  }
}
```

**Recommendation 5: Implement ML Component Orchestration.**
Effective orchestration—the coordinated management and execution of ML components—in dynamic product configurations requires an infrastructure that enables flexible model composition, state management between runs, and contextual integration. For this, we recommend:

- Use of modular ML pipelines, which allow integrating, monitoring, and scaling ML components in distributed environments.

- Intelligent orchestrators that dynamically adjust component activation according to contextual signals, business rules, or environmental conditions. Techniques such as context-aware scheduling can be applied.
- Functional decoupling of components, promoting a microservices-based architecture to facilitate model replacement, enhancement, or re-trainability without altering the overall configuration.
- Instrumentation for traceability and versioning: employ systems that record training data, parameters, results, and decisions made by each component to facilitate audits and optimization.

**Practical Implementation:** To operationalize this recommendation, SPL engineers must define a dedicated orchestration layer that governs the lifecycle, dependencies, and interactions of ML components. This orchestration must support declarative workflows, dynamic adaptation policies, and seamless integration with monitoring systems. The following schema defines a formal representation of such an orchestration-aware product configuration:

```
ProductConfiguration = {
  configuration_id: String,
  feature_binding: Map[Feature, ComponentBinding],
  workflow_specification: {
    component_graph: DirectedAcyclicGraph[Component, DataFlow],
    execution_constraints: Set[Constraint],
    quality_objectives: Map[QualityAttribute, Objective],
    resource_allocations: Map[Component, ResourceAllocation]
  }, adaptation_policies: {
    monitoring_configuration: MonitoringPolicy,
    replacement_triggers: Set[ReplacementTrigger],
    quality_negotiation: QualityNegotiationStrategy,
    performance_optimization: OptimizationPolicy
  }, validation_requirements: {
    functional_tests: Set[TestSpecification],
    performance_benchmarks: Set[BenchmarkTest],
    quality_assertions: Set[QualityAssertion],
    compliance_checks: Set[ComplianceCheck]
  }
}
```

This schema defines the structural and behavioral dimensions of a configurable product instance. Its modular design supports precise, verifiable, and adaptive configuration management across a wide range of variability. The key components are described below:

- **configuration_id**: A unique identifier assigned to the product configuration instance.

- **feature_binding**: A mapping between product features and their corresponding component implementations. This allows resolution of variability by specifying which components realize which features in a given configuration.

- **workflow_specification**: Captures the operational logic of the product.

    - **component_graph**: A directed acyclic graph (DAG) that defines the data flow and execution dependencies among software and ML components.
    - **execution_constraints**: A set of logical or resource-based constraints that govern component execution (e.g., timing, sequencing).

- ○ **quality_objectives**: Specifies target values for quality attributes, such as accuracy, latency, and energy consumption.
- ○ **resource_allocations**: Assigns computational resources (e.g., CPU, memory, GPU) to each component to ensure operational feasibility.

● **adaptation_policies**: Define the runtime behavior of the product under varying operational conditions:

- ○ **monitoring_configuration**: Indicates how system performance is monitored during execution.
- ○ **replacement_triggers**: Defines conditions under which components should be replaced.
- ○ **quality_negotiation**: Specifies strategies for balancing competing quality attributes under constraints.
- ○ **performance_optimization**: Policies for dynamically optimizing performance based on monitored feedback.

● **validation_requirements**: Ensures that configured products meet their intended goals and regulatory requirements:

- ○ **functional_tests**: Set of specifications for functional correctness.
- ○ **performance_benchmarks**: Benchmark tests that measure system performance under predefined workloads.
- ○ **quality_assertions**: Verifiable and testable statements specifying the quality attributes that a configured product is required to meet.
- ○ **compliance_checks**: Formal checks to ensure adherence to standards, certifications, or domain-specific regulations.

**E-commerce Example:** In an online retail SPL, a dynamic product configuration may include ML components for personalized recommendations, fraud detection, and sentiment analysis. The component bindings for each product instance can vary significantly based on factors such as the target audience, expected transaction volume, and specific regional compliance mandates.

**Recommendation 6: Implement Systematic ML Component Replacement Strategy.**

During product configuration, an automated strategy should be established to update or replace ML components when performance degradation is detected. This requires the definition of an intervention mechanism that is triggered when the performance metrics of an ML component fall below predefined thresholds. The mechanism must support replacing the underperforming component with one of several alternatives: another ML model, a traditional software component, or, if appropriate, the temporary exclusion of the affected functionality from the system's execution flow.

**Practical Implementation:** To operationalize this recommendation, SPL engineers must define a replacement strategy component associated with each ML-enabled domain component. This component is responsible for responding to degradation alerts issued by the monitoring system and executing the actions defined in the replacement policy. The structure of the replacement strategy component can be formally specified as follows:

```
MLComponentReplacementStrategy = {
  component_id: String,
  replacement_hierarchy: {
```

```
    primary_alternative: ComponentReference,
    secondary_alternatives: List[ComponentReference],
    fallback_strategy: FallbackApproach
  }
}
```

This specification defines the structure required to enable resilient and automated replacement mechanisms for ML components. The attributes are described below:

- **component_id**: Unique identifier of the ML component.
- **replacement_hierarchy**: Hierarchy of alternatives in case of model degradation.

    - **primary_alternative**: Component directly prepared to take over the current ML model.
    - **secondary_alternatives**: List of additional (less optimal) alternatives.
    - **fallback_strategy**: Emergency strategy to continue providing service with reduced capabilities (e.g., AllowAll, ConservativeRuleBasedBlocking, RuleBasedBlocking, ManualReview, GracefulShutdown).

**E-commerce Example:** In an online retail SPL, a replacement strategy can be defined for the sentiment analysis component using both traditional and ML-based alternative models. To ensure system resilience, if no alternative component meets the required quality thresholds, a predefined fallback strategy is triggered, such as temporarily deactivating the sentiment analysis feature from the process flow.

```
MLComponentReplacementStrategy: {
    component_id: "tc_001",
    replacement_hierarchy: {
      primary_alternative: {
        id: "cardiffnlp/twitter-roberta-base-sentiment-latest",
        type: "ml_model", reason: "Most compatible fine-tuned model"
      },
      secondary_alternatives: [{
          id: "distilbert-base-uncased-sentiment",
          type: "ml_model",
          reason: "Lightweight model for fallback"
        }, {
          id: "rule_based_sentiment_classifier_v1",
          type: "software_component",
          reason: "Legacy rules-based classifier for conservative estimation"
      }],
      fallback_strategy: {type: "RuleBasedBlocking" }
    }}
```

## 3.4. Dynamic Product Configuration

Incorporating ML components during product configuration adds substantial depth to the variability and intelligence of SPL. However, configuration decisions must balance multiple competing objectives, such as performance, cost, and reliability, often under shifting operational conditions.

**Recommendation 7: Establish Multi-Objective Configuration Optimization.**

To enhance the adaptability and performance of ML-enabled SPLs, it is essential to establish multi-objective configuration optimization mechanisms. This approach enables organizations to simultaneously evaluate and balance competing concerns, including accuracy, latency, resource consumption, interpretability, and ethical constraints. By leveraging advanced optimization techniques such as Pareto efficiency or evolutionary algorithms, teams can generate configuration sets that meet diverse stakeholder requirements without compromising system integrity. Implementing multi-objective optimization also promotes continuous improvement, enabling dynamic reconfiguration as environments evolve or model behaviors drift over time.

**Practical Implementation:** To operationalize multi-objective configuration optimization, it is first necessary to formalize a set of competing objectives, such as model accuracy, latency, resource utilization, interpretability, and compliance with ethical standards, into quantifiable metrics. The configuration space should encompass both system-level parameters and ML-specific settings, including hyperparameters and pipeline structures. The exploration of trade-offs across this space can be conducted using optimization techniques such as evolutionary algorithms (e.g., NSGA-II), Bayesian multi-objective methods, or Pareto-based analysis. Configurations are evaluated through simulation or benchmarking, producing Pareto-optimal sets that offer balanced solutions. These sets can be visualized or presented through decision-support interfaces to facilitate selection based on dynamic stakeholder priorities and preferences. Finally, integrating the optimization process within CI/CD pipelines ensures continuous reconfiguration in response to model drift or changing operational constraints.

### 3.5. Product Derivation and Validation

This phase considers the methodology for deriving specific products from a configurable architecture, detailing how optimization criteria and stakeholder requirements guide the selection process. It also describes the validation mechanisms employed to ensure that the resulting products meet expected standards of functionality, performance, and reliability prior to deployment.

**Recommendation 8: Implement validation and testing strategies specifically designed for ML-enhanced products.**

Validation and testing strategies should incorporate both functional and non-functional assessments, including unit and integration testing, model performance evaluation across diverse datasets, fairness audits, and resource utilization benchmarking. In addition, these strategies should extend to include ML-specific validation approaches, such as statistical performance validation, bias detection testing, adversarial robustness assessment, and long-term stability verification. It must also support automated validation pipelines integrated into CI/CD workflows, enabling continuous monitoring and the rapid detection of anomalies, drift, or compliance violations.

**Practical Implementation:** To implement this recommendation, the first step is to configure the derivation. This involves selecting binary features and determining the quality distributions for ML components. It is essential to establish optimization criteria and stakeholder requirements to guide the automated selection of ML components. This ensures that each product is customized to meet its use case requirements. Once a product is derived, the unit tests, integration tests, and non-functional requirement tests must be executed. The process is further enhanced with ML-specific validations such as bias detection, adversarial robustness assessments, and long-term stability verification to address the unique vulnerabilities of machine learning models. The overall goal is to ensure that the derived products meet all expected standards of functionality, performance, and reliability prior to deployment. Finally, to ensure long-term reliability, all of these validation strategies are integrated into automated CI/CD pipelines.

To conclude, the proposed framework's recommendations, which encompass the entire lifecycle of ML-enhanced SPLs, have been partially implemented within the **VariaMos web tool**. An excerpt from this implementation is illustrated in Figure 2. These recommendations are currently being applied to the development of two proof-of-concept SPL—an **e-commerce SPL** and a **text editor SPLs**—enabling us to validate the framework's practical effectiveness empirically.

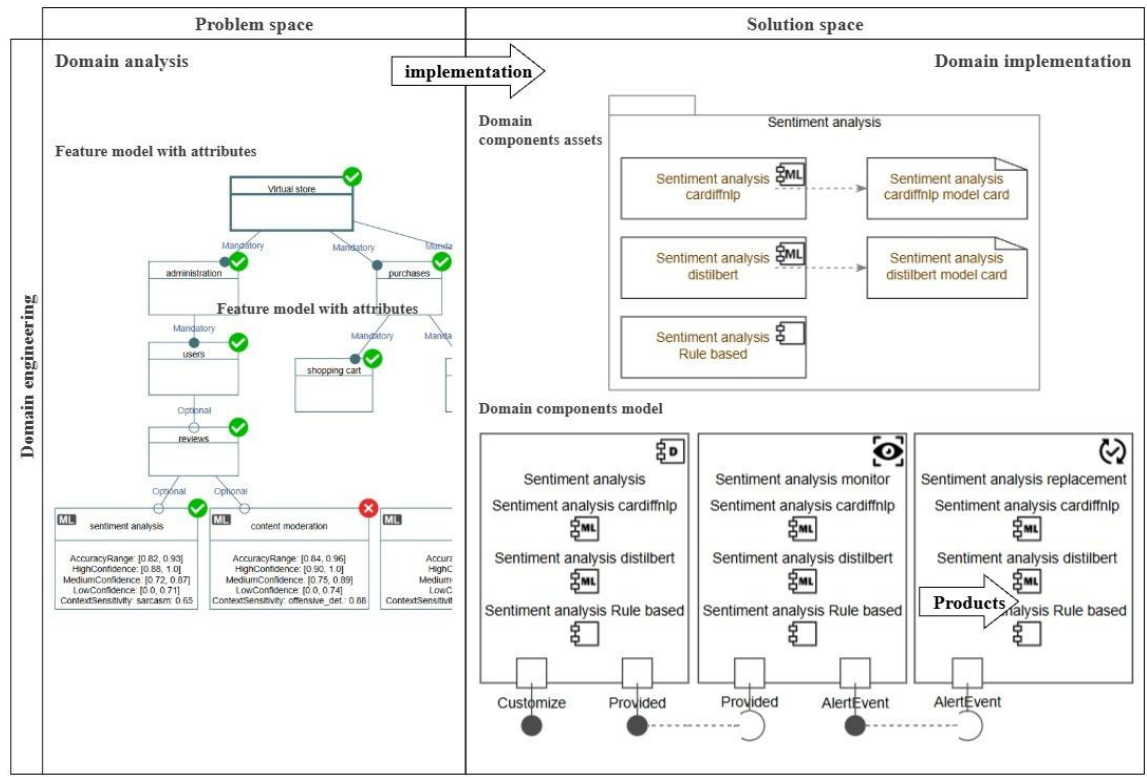

Figure 2. Partial implementation of ML-enhanced SPL Framework in VariaMos Tool

## 4. RELATED WORK

The intersection of SPL engineering and ML represents an emerging research area that builds upon established foundations in both domains. Traditional SPL engineering, formalized through seminal work by Clements, Mazo, and Pohl, respectively [7], [18], [19], has established comprehensive methodologies for systematic software reuse through domain engineering and application engineering processes. The Feature-Oriented Domain Analysis (FODA) approach introduced by Kang [8] and subsequent advances in variability management [7], [23] provide robust frameworks for managing product family complexity. However, these approaches fundamentally assume component determinism and behavioral predictability, creating significant gaps when dealing with probabilistic ML components.

Parallel developments in ML engineering have addressed the unique challenges of ML-enabled systems through comprehensive frameworks for technical debt management [22], engineering practices [24], and quality assurance approaches [25]. The emergence of systematic documentation practices, as exemplified by Model Cards [20] and behavioral testing frameworks [26], represents substantial progress in ML system engineering. Recent systematic reviews by Martínez-Fernández [4] and empirical studies by Nahar and Ribeiro, respectively [5], [26] have documented collaboration challenges and the complexity of requirements engineering specific to ML systems. Nevertheless, this body of work predominantly focuses on standalone ML systems or monolithic application contexts, with limited consideration of systematic reuse frameworks.

Architectural approaches for ML integration have evolved toward microservices-based patterns [27] and adaptive system frameworks [28], [29], while dynamic SPL research [7], [30], [31] has explored evolution and adaptation in product line contexts. However, existing approaches have not systematically addressed the unique requirements of ML components within SPL environments, including cross-product consistency management, shared component instance coordination, and the specific adaptation patterns required for probabilistic components subject to performance degradation and concept drift.

Current literature reveals critical limitations when applied to ML-enhanced SPL contexts. Traditional SPL methodologies assume behavioral predictability, which is incompatible with the probabilistic nature of ML components. In contrast, ML engineering approaches lack systematic frameworks for ensuring cross-product consistency and shared component management. Existing documentation frameworks do not provide reusability assessment mechanisms required for SPL component selection, and current adaptive system approaches do not address ML-specific degradation patterns and monitoring requirements.

This work addresses these fundamental gaps by providing the first comprehensive framework specifically designed to integrate ML components within SPLs, while preserving the benefits of systematic reuse. Unlike existing approaches that treat ML components as standalone services or apply ad-hoc integration patterns, our framework systematically extends established SPL methodologies with ML-specific concepts, including probabilistic feature modeling, degradation-aware component characterization, adaptive architectural patterns, and dynamic configuration optimization. The framework proposed in this paper provides concrete specifications, including formal orchestration languages (MCOSL), systematic monitoring frameworks, and multi-objective optimization approaches, enabling practitioners to maintain an engineering discipline and leverage systematic reuse advantages while effectively utilizing ML capabilities across products derived from product lines.

## 5. CONCLUSIONS AND FUTURE WORK

The integration of ML components into SPLs presents new challenges that traditional modeling techniques are not equipped to address. By addressing the variability and uncertainty inherent in ML components, this approach lays the groundwork for bridging the gap between SPLE and AI-based software development.

In this paper, we propose a framework that supports the inclusion of ML components in SPLs, facilitating systematic reuse, customization, and evolution. Our contribution consists of a specification-oriented approach that guides the integration of ML-based functionalities into SPLs, along with a set of recommendations and practical implementations. The framework is structured around five interconnected phases that encompass the entire lifecycle of ML-enhanced SPLs: ML-aware domain analysis, Adaptive architecture design, ML-aware domain implementation, Dynamic product configuration, and Product derivation and validation of its resulting products.

The proposed framework has been partially implemented in the **VariaMos** tool (https://variamos.com/). This web-based tool utilizes microservices to enable the specification of product lines through a multi-language modeling approach and the reasoning on these products and product lines. Initial empirical findings, obtained by applying these recommendations to two distinct SPL—an e-commerce SPL and a text editor SPL—suggest that this comprehensive documentation approach facilitates informed decision-making across the entire ML component lifecycle. This process spans from initial model selection to deployment and ongoing monitoring. Furthermore, Model Cards support regulatory compliance and risk management by providing auditable documentation of model characteristics and decision rationale, which contributes to the development of more accountable and trustworthy ML systems. Although these preliminary results are promising, further experimentation and implementation improvements are needed to fully assess the actual value and impact of this proposal in production environments. Future work involves evaluating the proposed strategy in real-world industrial domains, including a detailed cost-benefit analysis, extending the capabilities of the VariaMos tool, and exploring its applicability to AI components beyond ML. Furthermore, in the educational context, we plan to move beyond the individual courses on "Software Engineering for ML-enabled Systems" and "Software Product Line Engineering" currently offered at each institution by developing a joint course in which students collaboratively design SPLs to address real-world problems.

**ACKNOWLEDGMENTS**

This research was supported by the University of Antioquia, Colombia, through the Committee for the development of research – CODI (PRV2022-52951), and the ENSTA, France.

## AUTHORS

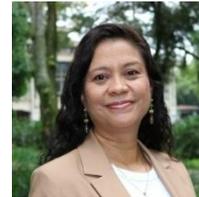

**Luz-Viviana Cobaleda** is an Associate Professor at the University of Antioquia, Colombia. She received her Ph.D. in Electronic Engineering with an emphasis on Software Engineering, her Master's degree in Engineering, and her B.Sc. in Systems Engineering, all from the University of Antioquia, as well as a Specialization in Software Engineering from EAFIT University, Colombia.Her research interests include software engineering for AI/ML-enabled systems (SE4AI), as well as software engineering methods and techniques—such as model-driven development, software design and specification, and adaptive and personalized systems—and requirements engineering. She has published papers in international conferences and journals and actively participates in collaborative research projects in software engineering, including adaptive and personalized software systems, at the University of Antioquia.

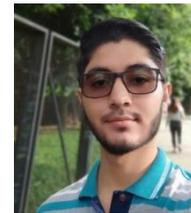

**Julian Carvajal** is a Colombian software engineer in training and a Systems Engineering student at the University of Antioquia (Colombia). He has professional experience as a software developer, with a particular focus on building educational video games for preschool children and contributing to research-driven software projects. He has also collaborated in the VariaMos project, developing a tool to help users define the graphical representation of modeling languages. He is currently working on SignAI UdeA, an initiative that leverages artificial intelligence to facilitate communication between deaf and hearing people through the recognition of Colombian Sign Language (LSC). His research interests include software engineering for artificial intelligence (SE4AI), particularly the integration of AI components into software systems.

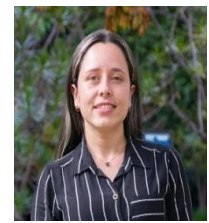

**Paola Vallejo** is a Systems Engineer graduated from Universidad EAFIT in 2012. She got her Master degree (Human Computer Centered Systems) at École Nationale d'Ingénieurs de Brest - France in 2012. She received the Ph. D. degree in Computer Science from Université de Bretagne Occidentale - France in 2015. She is currently a full professor at Universidad EAFIT. She has also collaborated in the VariaMos working group. Her research interests include the reuse of software components, model-driven engineering, requirements engineering, software architecture, and human–computer interactions.

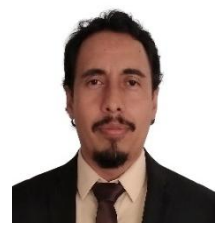

**Andrés Orlando López Henao** is a Colombian systems engineer. He graduated in 2013 with a degree in Systems Engineering from the University of Antioquia (Colombia) and in 2018 obtained a Master's in Engineering from EAFIT University (Colombia). He is currently pursuing a joint Ph.D. in Sciences pour l'ingénieur et le numérique at the École Nationale Supérieure de Techniques Avancées – ENSTA (France) and in Electronic and Computer Engineering at the University of Antioquia (Colombia). His research interests focus on software product lines, artificial intelligence, and software engineering.

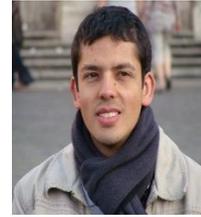

**Raúl Mazo** is a Franco-Colombian engineer who received his Engineering degree in Informatics from the University of Antioquia (Colombia) in 2005, and later earned an M.S. in Information Systems, a Ph.D. in Computer Science, and the Habilitation à Diriger des Recherches (HDR) from the University Panthéon-Sorbonne (France) in 2008, 2011, and 2018, respectively. He is currently a Full Professor at the École Nationale Supérieure de Techniques Avancées (ENSTA). Prior to this, he served as an Associate Professor at Panthéon-Sorbonne University and worked as a software developer and telecommunications engineer in small and medium-sized enterprises.

His research and teaching interests include model-driven engineering, requirements engineering, variability management, software & systems architecture, and artificial reasoning. He leads the VariaMos working group and tool, through which he has contributed to numerous national, European, and intercontinental research initiatives.